\documentclass[epj]{webofc}
\usepackage[varg]{txfonts}   % Web of Conferences font
\woctitle{MESON2016 - the 14$^\textrm{th}$ International Workshop on Meson 
Production, Properties and Interaction}
\usepackage{latexsym}
\usepackage{amsmath}
\usepackage{amssymb}
\usepackage{amsfonts}

\usepackage{color}

\usepackage{supertabular} 
\usepackage{placeins}
\usepackage{epsfig}
\usepackage{graphicx}

\begin{document}
\selectlanguage{english}
\title{Molecular components in $\mathbf{D_{s0}^{\ast}(2317)}$ and 
$\mathbf{D_{s1}(2460)}$ mesons}

% insert email only for speaker/presenter
\author{
Pablo~G.~Ortega\inst{1}\footnote{Currently at Instituto de F\'isica Corpuscular 
(IFIC), CSIC-Universidad de Valencia, E-46071 Valencia, Spain
} 
\and
Jorge~Segovia\inst{2}\fnsep\thanks{speaker, \email{jorge.segovia@tum.de}}
\and
David~R.~Entem\inst{3}
\and
Francisco Fern\'andez\inst{3}
% comment out the next line if not needed
%       \\for the XXXXX Collaboration
}

\institute{CERN (European Organization for Nuclear Research), CH-1211 Geneva, 
Switzerland
\and
Technische Universit\"at M\"unchen, D-85748 Garching bei M\"unchen, Germany
\and
Universidad de Salamanca, E-37008 Salamanca, Spain
}

\abstract{%Do not break line here!
Different experiments have confirmed that the $D_{s0}^{\ast}(2317)$ and 
$D_{s1}(2460)$ mesons are very narrow states located, respectively, below the 
$DK$ and $D^{\ast}K$ thresholds. This is markedly in contrast with the 
expectations of naive quark models and heavy quark symmetry. We address the 
mass shifts of the $c\bar{s}$ ground states with quantum numbers $J^{P}=0^{+}$ 
($D_{s0}^{\ast}(2317)$) and $J^{P}=1^{+}$ ($D_{s1}(2460)$) using a 
nonrelativistic constituent quark model in which quark-antiquark and 
meson-meson degrees of freedom are incorporated. The quark model has been 
applied to a wide range of hadronic observables and thus the model parameters 
are completely constrained. We observe that the coupling of the $0^{+}$ 
$(1^{+})$ meson sector to the $DK$ $(D^{\ast}K)$ threshold is a key feature in 
lowering the masses of the corresponding $D_{s0}^{\ast}(2317)$ and 
$D_{s1}(2460)$ states predicted by the naive quark model, but also in 
describing the $D_{s1}(2536)$ meson as the $1^{+}$ state of the 
$j_{q}^{P}=3/2^{+}$ doublet predicted by heavy quark symmetry and thus 
reproducing its strong decay properties. Two features of our formalism cannot 
be address nowadays by other approaches: the coupling of the $D$-wave 
$D^{\ast}K$ threshold in the $J^{P}=1^{+}$ $c\bar{s}$ channel and the 
computation of the probabilities associated with different Fock components in 
the physical state.
}
\maketitle
%%%%%%%%%%%%%%%%%%%%%%%%%%%%%%%%%%%%%%%%%%%%%%%%%%%%%%%%%%%%%%%%%%%%%%%%%%%%%%%
%%%%%%%%%%%%%%%%%%%%%%%%%%%%%%%%%%%%%%%%%%%%%%%%%%%%%%%%%%%%%%%%%%%%%%%%%%%%%%%

\vspace*{-0.50cm}
\section{Introduction}
\label{sec:Introduction}

Prior to the discovery in $2003$ of the $D_{s0}^{\ast}(2317)$ 
$(J^{P}=0^{+})$~\cite{Aubert:2003fg} and $D_{s1}(2460)$ 
$(1^{+})$~\cite{Besson:2003cp} resonances, the heavy-light meson sectors were 
reasonably well understood in the $m_{Q}\to\infty$ limit. In such a limit, 
heavy quark symmetry (HQS) holds~\cite{Isgur:1991wq}. The heavy quark acts as a 
static color source, its spin $s_{Q}$ is decoupled from the total angular 
momentum of the light quark $j_{q}$ and they are separately conserved. Then, 
the heavy-light mesons can be organized in doublets, each one corresponding to 
a particular value of $j_{q}$ and parity. For the lowest $P$-wave 
charmed-strange mesons, HQS predicts two doublets which are labeled by 
$j_q^P=1/2^{+}$ with $J^P=0^{+},\,1^{+}$ and $j_q^P=3/2^+$ with 
$J^P=1^+,\,2^+$. Moreover, the strong decays of the $D_{sJ}\,(j_q=3/2)$ proceed 
only through $D$-waves, while the $D_{sJ}\,(j_q=1/2)$ decays happen only 
through $S$-waves~\cite{Isgur:1991wq}. The $D$-wave decay is suppressed by the 
barrier factor which behaves as $q^{2L+1}$ where $q$ is the relative momentum of 
the two decaying mesons. Therefore, states decaying through $D$-waves are 
expected to be narrower than those decaying via $S$-waves. 

The $D_{s0}^{\ast}(2317)$ and $D_{s1}(2460)$ mesons are considered to be the 
members of the $j_q^{P}=1/2^{+}$ doublet and thus being almost degenerated and 
broad due to its $S$-wave decay. However, neither experimental values of their 
masses nor their empirical widths accommodate into the theoretical 
expectations. These results led to many theoretical speculations about the 
nature of these resonances ranging from conventional $c\bar{s}$ 
states~\cite{Fayyazuddin:2003aa, Lakhina:2006fy} to molecular or compact 
tetraquark interpretations~\cite{Barnes:2003dj, Lipkin:2003zk, Bicudo:2004dx, 
Gamermann:2006nm, Gamermann:2007fi, Torres:2014vna, Guo:2015dha}. 

Certainly quark models predict $c\bar s$ ground states with quantum numbers 
$J^{P}=0^{+}$ and $1^{+}$ that do not fit the experimental data. As the 
predictions of the quark models are roughly reasonable for other states in the 
charmed-strange sector~\cite{Segovia:2015dia}, one must expect that the 
$D_{s0}^{\ast}(2317)$ and $D_{s1}(2460)$ resonances should be modifications of 
the genuine $c\bar s$ states rather than new states out of the systematics of 
the quark model. On this respect, particularly relevant was the 
suggestion~\cite{vanBeveren:2003kd, vanBeveren:2003jv} that the coupling of the 
$J^{P}=0^{+}$ $(1^{+})$ $c\bar{s}$ state to the $DK$ $(D^{\ast}K)$ threshold 
plays an important dynamical role in lowering the bare mass to the observed 
value. Moreover, in a recent lattice study of the $D_{s0}^{\ast}(2317)$ and 
$D_{s1}(2460)$ mesons~\cite{Lang:2014yfa}, good agreement with the experimental 
mass was found when operators for $D^{(\ast)}K$ scattering states are included.

In this contribution to the proceedings we present the work performed in 
Ref.~\cite{Ortega:2016mms}\footnote{All the details about the computation and 
the theoretical framework can be found in Ref.~\cite{Ortega:2016mms} and 
references therein.}. Therein, we study the low-lying $P$-wave 
charmed-strange mesons using a nonrelativistic constituent quark model in which 
quark-antiquark and meson-meson degrees of freedom are incorporated. The 
constituent quark model (CQM) was proposed in Ref.~\cite{Vijande:2004he} (see 
references~\cite{Valcarce:2005em} and~\cite{Segovia:2013wma} for reviews). 
This model successfully describes hadron phenomenology and hadronic 
reactions and has been recently applied to mesons containing heavy quarks (see, 
for instance, Refs.~\cite{Ortega:2010qq, Segovia:2011zza, Segovia:2013kg, 
Segovia:2013sxa, Segovia:2014mca}).

%%%%%%%%%%%%%%%%%%%%%%%%%%%%%%%%%%%%%%%%%%%%%%%%%%%%%%%%%%%%%%%%%%%%%%%%%%%%%%%
%%%%%%%%%%%%%%%%%%%%%%%%%%%%%%%%%%%%%%%%%%%%%%%%%%%%%%%%%%%%%%%%%%%%%%%%%%%%%%%

\begin{figure}[!t]
\centering
\includegraphics[height=0.225\textheight, width=0.90\textwidth,clip]
{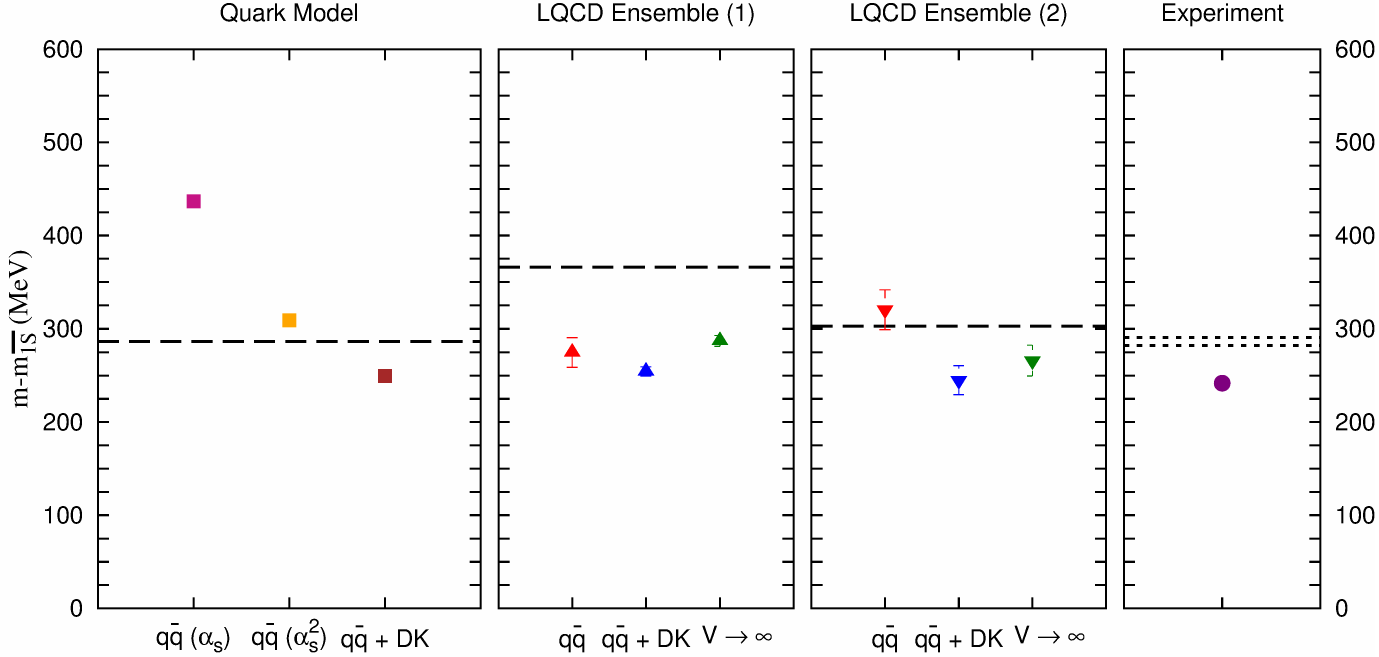}
\caption{\label{fig:Ds2317} Energy levels from constituent quark model (CQM), 
from Lattice QCD~\cite{Lang:2014yfa} using Ensemble~(1) and Ensemble~(2), and 
from experiment~\cite{Agashe:2014kda}. We show, for CQM results, the 
quark-antiquark value taking into account the OGE potential $(\alpha_{s})$, 
including its one-loop corrections $(\alpha_{s}^{2})$ and coupling with the 
$DK$ threshold. For the lattice QCD results, in each ensemble, we show values 
with just a $q\bar{q}$ interpolator basis and with a combined basis of 
$q\bar{q}$ and $DK$ interpolating fields. The value of the bound 
$D_{s0}^{\ast}(2317)$ state position in the infinite volume limit $V\to \infty$ 
is obtained by an analytical continuation of the scattering amplitude combined 
with L\"uscher's finite volume method. The dashed lines represent the threshold 
for $DK$ in each approach and the dotted lines are the thresholds for 
$D^{0}K^{+}$ and $D^{+}K^{0}$ in experiment. \vspace*{-0.60cm}}
\end{figure}

\vspace*{-0.30cm}
\section{Results for the $D_{s0}^{\ast}(2317)$ meson} 
\label{sec:0p} 

Figure~\ref{fig:Ds2317} compares our results for the $D_{s0}^{\ast}(2317)$ 
meson with the lattice QCD study of Ref.~\cite{Lang:2014yfa} and with 
experiment~\cite{Agashe:2014kda}. Instead of the $D_{s0}^{\ast}(2317)$ mass 
itself, following the lattice study, we compare the values of 
$m_{D_{s0}^{\ast}(2317)} - m_{\overline{1S}}$, where $m_{\overline{1S}} = 1/4 
(m_{D_{s}} + 3m_{D_{s}^{\ast}})$ is the spin-averaged ground state mass.

The mass of the $D_{s0}^{\ast}(2317)$ state obtained using the naive quark 
model and without the $1$-loop corrections to the one-gluon exchange (OGE) 
potential is much higher than the experimental value. In this case, the 
$m_{D_{s0}^{\ast}(2317)} - m_{\overline{1S}}=437\,{\rm MeV}$ is almost twice the 
empirical figure. The mass associated to the $D_{s0}^{\ast}(2317)$ state is very 
sensitive to the $\alpha_{s}^{2}$-corrections of the OGE potential. This effect 
brings down the $m_{D_{s0}^{\ast}(2317)} - m_{\overline{1S}}$ splitting to 
$309\,{\rm MeV}$, which is now only $30\%$ higher than the experimental value. 
However, as one can see in Fig.~\ref{fig:Ds2317}, the hypothetical 
$D_{s0}^{\ast}(2317)$ state would be above the $DK$ threshold and thus would 
decay into this final channel in an $S$-wave making the state wider than the 
observed one. The mass-shift due to the $\alpha_{s}^{2}$-corrections allows that 
the $0^{+}$ state be close to the $DK$ threshold. This makes the $DK$ coupling a 
relevant dynamical mechanism in the formation of the $D_{s0}^{\ast}(2317)$ bound 
state. When we couple the $0^{+}$ $c\bar{s}$ ground state with the $DK$ 
threshold, the splitting $m_{D_{s0}^{\ast}(2317)} - 
m_{\overline{1S}}=249.6\,{\rm MeV}$ is in good agreement with experiment. 
Regarding the probabilities of the different Fock components in the physical 
state, we obtain $66\%$ for $q\bar q$ and $34\%$ for $DK$ reflecting that the 
$D_{s0}^{\ast}(2317)$ meson is mostly of quark-antiquark nature in our approach.

%%%%%%%%%%%%%%%%%%%%%%%%%%%%%%%%%%%%%%%%%%%%%%%%%%%%%%%%%%%%%%%%%%%%%%%%%%%%%%%
%%%%%%%%%%%%%%%%%%%%%%%%%%%%%%%%%%%%%%%%%%%%%%%%%%%%%%%%%%%%%%%%%%%%%%%%%%%%%%%

\begin{figure}[!t]
\centering
\includegraphics[height=0.225\textheight,width=0.90\textwidth,clip]
{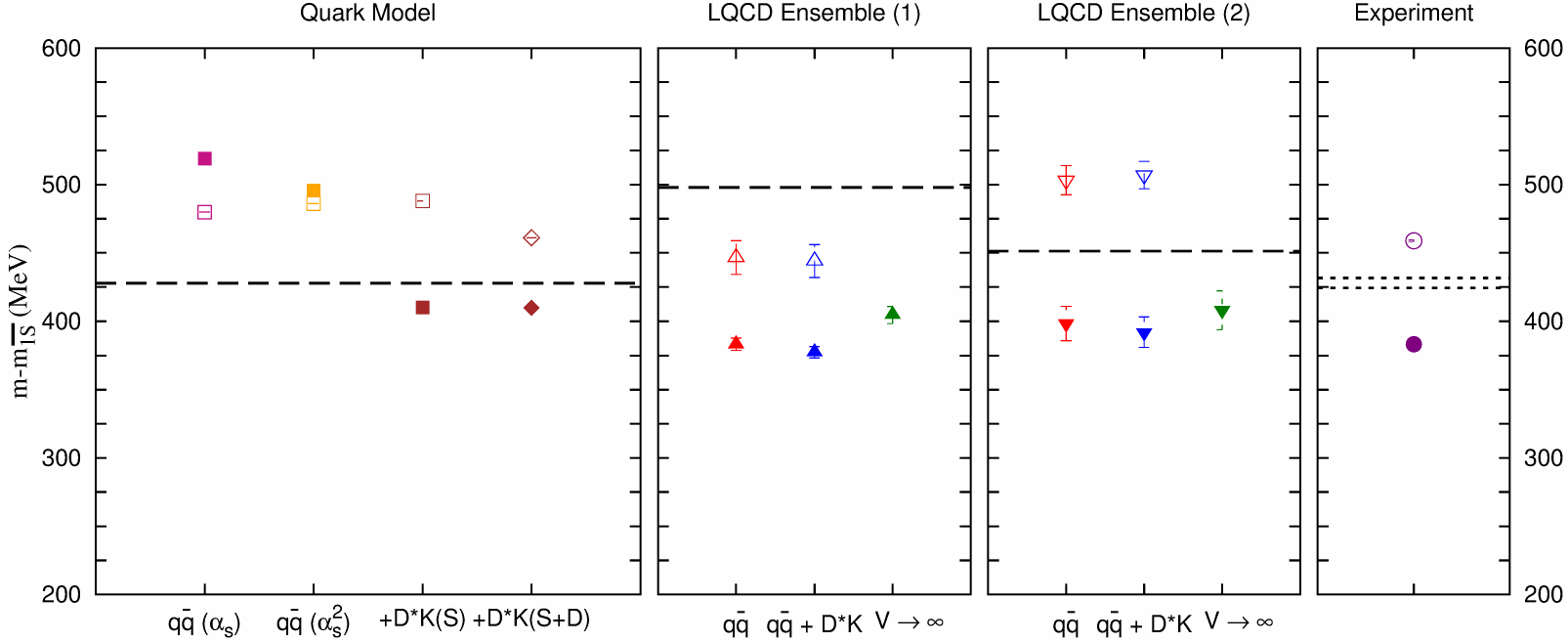}
\caption{\label{fig:Ds24602536} Energy levels from constituent quark model 
(CQM), from Lattice QCD~\cite{Lang:2014yfa} using Ensemble~(1) and 
Ensemble~(2), and from experiment~\cite{Agashe:2014kda}. We show, for CQM 
results, the quark-antiquark value taking into account the OGE potential 
$(\alpha_{s})$, including its one-loop corrections $(\alpha_{s}^{2})$ 
and coupling with the $D^{\ast}K$ threshold in $S$- and $D$-wave. For the 
lattice QCD results, in each case, we show values with just a $q\bar{q}$ 
interpolator basis and with a combined basis of $q\bar{q}$ and $D^{\ast}K$ 
interpolating fields. Remember that in the lattice QCD computations the 
$D^{\ast}K$ threshold is coupled only in an $S$-wave. The value of the bound 
$D_{s1}(2460)$ state position in the infinite volume limit $V\to \infty$ is 
obtained by an analytical continuation of the scattering amplitude combined 
with L\"uscher's finite volume method. This method has not been used for the 
$D_{s1}(2536)$ meson. The dashed lines represent the threshold for $D^{\ast}K$ 
in each approach and the dotted lines are the thresholds for $D^{\ast0}K^{+}$ 
and $D^{\ast+}K^{0}$ in experiment. \vspace*{-0.60cm}}
\end{figure}

\vspace*{-0.30cm}
\section{Results for the $D_{s1}(2460)$ and $D_{s1}(2536)$ mesons}
\label{sec:1p}

Figure~\ref{fig:Ds24602536} compares our results for the $m_{D_{s1}} - 
m_{\overline{1S}}$ mass splitting of the first two $J^{P}=1^{+}$ 
charmed-strange states with the lattice QCD study of Ref.~\cite{Lang:2014yfa} 
and with experiment~\cite{Agashe:2014kda}.

The naive quark model predicts that the states corresponding to the 
$D_{s1}(2460)$ and $D_{s1}(2536)$ mesons are almost degenerated, with masses 
close to the experimentally observed mass of the $D_{s1}(2536)$. The inclusion 
of the $1$-loop corrections to the OGE potential does not improve the 
situation, making the splitting between the two states even smaller. Following 
lattice criteria, we couple first the $D^{\ast}K$ threshold in an $S$-wave with 
the two $1^{+}$ $c\bar{s}$ states. One can see in Fig.~\ref{fig:Ds24602536} 
that the state associated with the $D_{s1}(2460)$ meson goes down in the 
spectrum and it is located below $D^{\ast}K$ threshold with a mass compatible 
with the experimental value. The state associated with the $D_{s1}(2536)$ meson 
is almost insensitive to this coupling because it is the $J^{P}=1^{+}$ member 
of the $j_{q}=3/2$ doublet predicted by HQS and thus it couples mostly in a 
$D$-wave to the $D^{\ast}K$ threshold. Lattice QCD has not yet computed the 
coupling in $D$-wave of the $D^{\ast}K$ threshold with the $1^{+}$ $c\bar{s}$ 
sector. This coupling is trivially implemented in our approach. The state 
associated with the $D_{s1}(2460)$ meson experience a very small modification 
because it is almost the $|1/2,1^{+} \!\!\left.\right\rangle$ eigenstate of 
HQS, whereas the state associated with $D_{s1}(2536)$ meson suffers a moderate 
mass-shift approaching to the experimental value. 

When the $D^{\ast}K$ threshold is coupled, the meson-meson component is around 
$50\%$ for both $D_{s1}(2460)$ and $D_{s1}(2536)$ mesons. It is also relevant 
to realize that the quark-antiquark component in the wave function of the 
$D_{s1}(2536)$ meson holds quite well the $^{1}P_{1}$ and $^{3}P_{1}$ 
composition predicted by HQS, which is crucial in order to have a very narrow 
state and describe well its decay properties.

%%%%%%%%%%%%%%%%%%%%%%%%%%%%%%%%%%%%%%%%%%%%%%%%%%%%%%%%%%%%%%%%%%%%%%%%%%%%%%%
%%%%%%%%%%%%%%%%%%%%%%%%%%%%%%%%%%%%%%%%%%%%%%%%%%%%%%%%%%%%%%%%%%%%%%%%%%%%%%%

\vspace*{-0.30cm}
\section{Summary}
\label{sec:summary}

We have performed a coupled-channel computation taking into account the 
$D_{s0}^{\ast}(2317)$, $D_{s1}(2460)$ and $D_{s1}(2536)$ mesons and the $DK$ 
and $D^{\ast}K$ thresholds within the framework of a constituent quark model. 
Our method allows to introduce the coupling with the $D$-wave $D^{\ast}K$ 
channel and the computation of the probabilities associated with the different 
Fock components of the physical state.
% , features  which cannot be addressed nowadays by any other theoretical 
% formalism.

%%%%%%%%%%%%%%%%%%%%%%%%%%%%%%%%%%%%%%%%%%%%%%%%%%%%%%%%%%%%%%%%%%%%%%%%%%%%%%%
%%%%%%%%%%%%%%%%%%%%%%%%%%%%%%%%%%%%%%%%%%%%%%%%%%%%%%%%%%%%%%%%%%%%%%%%%%%%%%%

\vspace*{0.20cm}
\begin{acknowledgement}
This work has been partially funded by Ministerio de Ciencia y Tecnolog\'\i a 
under Contract no. FPA2013-47443-C2-2-P, by the Spanish Excellence Network 
on Hadronic Physics FIS2014-57026-REDT, and by the Junta de Castilla y Le\'on 
under Contract no. SA041U16. P.G.O. acknowledges the financial support from the 
Spanish Ministerio de Economia y Competitividad and European FEDER funds under 
the contract no. FIS2014-51948-C2-1-P. J.S. acknowledges the financial support 
from Alexander von Humboldt Foundation.
\end{acknowledgement}
\vspace*{-0.25cm}
% BibTeX or Biber users please use (the style is already called in the class, 
%ensure that the "woc.bst" style is in your local directory)
\bibliography{JorgeSegovia_MESON2016}
%
% Non-BibTeX users please use
%
% \begin{thebibliography}{00}
% %
% % and use \bibitem to create references.
% %
% \bibitem{RefJ}
% % Format for Journal Reference
% F.~Author~\textit{et al.}, Journal \textbf{Volume}, page numbers (year) {\em{no trailing dot!}}
% % Format for books
% \bibitem{RefB}
% Book Author, \textit{Book title} (Publisher, place, year) page numbers {\em{no trailing dot!}}
% % etc
% \end{thebibliography}

\end{document}